\documentclass{article}
\usepackage{graphicx}
\def\be{\begin{equation}}
\def\ee{\end{equation}}
\def\bea{\begin{eqnarray}}
\def\eea{\end{eqnarray}}
\def\Journal#1#2#3#4{{#1} {\bf #2}, #3 (#4)}

\def\NCA{\em Nuovo Cimento}
\def\NIM{\em Nucl. Instrum. Methods}
\def\NIMA{{\em Nucl. Instrum. Methods} A}
\def\NPB{{\em Nucl. Phys.} B}
\def\PLB{{\em Phys. Lett.}  B}
\def\PRL{\em Phys. Rev. Lett.}
\def\PRD{{\em Phys. Rev.} D}
\def\ZPC{{\em Z. Phys.} C}

\begin{document}
\vspace*{3cm}
\begin{center}
\textbf{RECENT ADVANCES IN ODDERON PHYSICS}\\
\vspace*{1cm}
Basarab NICOLESCU
\end{center}

\begin{center}
LPNHE\footnote{Unit\'e de Recherche des Universit\'es Paris 6 et 
        Paris 7, Associ\'ee au CNRS} - LPTPE, Universit\'e Pierre et Marie 
        Curie, 4, Place Jussieu,\\ 75252 Paris Cedex 05, France\\
        E-mail: nicolesc@ipno.in2p3.fr
\end{center}

\vspace*{1cm}

\begin{center}
Talk at International Conference on Elastic and Diffractive Scattering
 (VIIIth EDS Blois Workshop), Protvino, Russia, June 28 - July 2, 1999\\
(to be published in the Proceedings of this Conference)
\end{center}

\vspace*{3cm}

\noindent Abstract

We present in this talk the phenomenological and theoretical 
advances in Odderon physics which occured since the last EDS Blois 
Workshop (Seoul, 1997).

\vfill

\noindent LPNHE 99-06 \hfill November 1999

\newpage

\section{Introduction}
The elastic and diffractive scattering is controlled by two 
singularities located near $J=1$ in the complex $J$-plane : the 
Pomeron\cite{Pom58}, in the even-under-crossing amplitude $F_{+}$ and 
the Odderon\cite{Lu73}, in the odd-under-crossing amplitude $F_{-}$.

The Odderon idea was longtime forced to stay in the Purgatory, because 
it contradicted the belief, founded on the dominant approach of the 
70's - the Regge-pole model, that $F_{-}$ is dominated by singularities 
located near $J=1/2$ ($\rho$ and $\omega$ Regge poles and their cuts). 
In spite of its rediscovery in QCD in the 80's\cite{Ba80}, and of its 
solid theoretical foundation in the framework of asymptotic 
theorems\cite{Lu73},\cite{Jo75} and derivative relations\cite{Ka75}, 
the Odderon continued to be considered as an heretical concept. As 
late as in October 1990, here, in Dubna, Andr\'e Martin did not 
hesitate to associate the words "revolution" and "Odderon"\cite{Ma90}.
The complete theoretical legitimacy of the Odderon came only in the 
last few years from the calculation of the Odderon intercept in QCD. 
In the first part of my talk I will discuss the last results in this 
field.

From the phenomenological point of view, the interest clearly shifted 
from the somewhat biased study of Odderon effects in $\bar pp$ and 
$pp$ scattering towards HERA Odderon physics. The surprisingly rich 
activity in Odderon phenomenology in the last two years will be 
described, before drawing conclusions, in the second part of the talk.

\section{Calculation of the Odderon intercept in QCD}
The most important recent result is, beyond any doubt, the discovery 
by Janik and Wosiek\cite{Ja99} and, independently, by 
Lipatov\cite{Li97}, of an exact solution for the Odderon intercept in 
LLA.

In QCD the Odderon is a C-odd
state of 3 reggeized gluons which interact pairwise with a well-defined
potential (Fig. 1). The problem is to find an operator  $\hat{q}_3$,
\begin{equation}
                \hat{q}_3^2 = - r_{12}^2 r_{23}^2 r_{31}^2 p_1^2 p_2^2 p_3^2 , 
                \label{eq:br1}
\end{equation}
which commutes with the Odderon hamiltonian H,
\begin{equation}
        [ \hat{q}_3^2,H ] = 0
        \label{eq:br2}
\end{equation}
and has a much simpler form than H. 

The exact solution of the problem is formulated in terms of the 
eigenequation
\be
\hat{q}_3 f =q_3 f
\label{eq:br3}
\ee
where
\bea
f &=& \left(\frac {\rho_{12} \rho_{13} \rho_{23}} 
{\rho^2_{10} \rho^2_{20} \rho^2_{30}}\right)^\mu \Phi(z),   
\label{eq:br4}
\eea
with $\mu=h/3$, h being the conformal weight.
The function $\Phi$  
satisfies a third order linear equation
\bea
a(z) {d^3 \over d z^3}\Phi(z) + b(z){d^2\over d z^2} \Phi(z) 
+c(z){d\over dz}\Phi(z) \nonumber \\ \noalign{\vskip 3pt} + d(z) \Phi(z)=0,
\label{eq:br5}
\eea
where
\be
a(z)=z^3 (1-z)^3,
\label{eq:br6}
\ee

\be
b(z)=2 z^2 (1-z)^2 (1-2z),
\label{eq:br7}
\ee

\be
c(z) = z(z-1)\left(z(z-1)(3\mu+2)(\mu-1)+3\mu^2-\mu\right),
\label{eq:br8}
\ee

\be
d(z)=\mu^2 (1-\mu)(z+1)(z-2)(2z-1) -iq_3 z(1-z),
\label{eq:br9}
\ee 

\be
z=\frac {\rho_{12} \rho_{30}} {\rho_{10} \rho_{32}},
\rho_k=x_k+iy_k (k=1,2,3), \rho_{ij}=\rho_i - \rho_j.   
\label{eq:jw3}
\ee

In Ref. 7, the numerical solution
\be
q_3=-0.20526\,i
\label{eq:br10}
\ee
was found, corresponding to the Odderon energy
\be
\epsilon=0.16478.
\label{eq:br11}
\ee

The relation between the Odderon energy $\epsilon$
and the Odderon intercept $\alpha_O(0)$
is given by the equation
\be
\alpha_O(0) = 1-(9\alpha_s/2\pi)\epsilon.
\label{eq:br12}
\ee

For realistic values of $\alpha_s$  ($\alpha_s \simeq 0.19$)
one gets

\be
\alpha_O(0) = 0.94.
\label{eq:br13}
\ee

In collaboration with M.A. Braun and P. Gauron we recently performed 
a direct calculation of the lower bound for the Odderon 
intercept\cite{Br99} in the framework of the variational approach we 
formulated earlier in collaboration with L. Lipatov\cite{Ga91}. In 
this variational approach the Odderon energy is defined as
\be
\epsilon = E/D
\label{eq:br14}
\ee

In Eq.(\ref{eq:br14}) D is a normalization constant and the energy 
functional is given by
\be
E=\sum_{n=-\infty}^{\infty}\int_{-\infty}^{\infty}d\nu\epsilon_{n}(\nu)
|\alpha_{n}(\nu)|^{2},
\label{eq:br15}
\ee 
where
\be
\epsilon_{n}(\nu)=2 Re[\psi(\frac{1+|n|}{2}+i\nu)-\psi(1)],
\label{eq:br16}
\ee 

\bea
\alpha_{n}(\nu) &=& \int_{0}^{\infty} dr r^{-2-2i\nu}\int_{0}^{2\pi} d\phi  
e^{-in\phi} \nonumber \\ \noalign{\vskip 3pt} &&{}
\left(i\nu+\frac{n+1}{2}+re^{i\phi}(h-i\nu-\frac{n-1}{2})\right)
(i\nu-\frac{n-1}{2}) \nonumber \\ \noalign{\vskip 3pt} &&{}
(-\tilde{h}+i\nu-\frac{n-1}{2}) Z(r,\phi),
\label{eq:br17}
\eea

\bea
h &=& 1/2+n/2-i\nu, \tilde{h} = 1/2-n/2+i\nu, \nonumber \\ 
 \noalign{\vskip 3pt} &&{} - \infty<\nu<\infty, n=\ldots ,-1,0,1, \ldots ,
\label{eq:br18}
\eea

\be
Z(z,z^*)=|z(1-z)|^{2h/3}\Psi(z,z^*).
\label{eq:br19}
\ee

The function $\Psi$
in Eq. (\ref{eq:br19}) is invariant under the transformations 
$z\rightarrow 1-z$ and $z\rightarrow 1/z$
(Bose symmetry in the 3 gluons).

The following trial functions $\Psi$
were used in Ref. 9 :

\be
\Psi = \sum_{k=1}^{N-N_1} c_k a^{k/2-1/6}
+\sum_{k=1}^{N_1} d_k a^{k-1/6}\ln a,
\label{eq:br20}
\ee

where

\be
a=\frac{r^2r_1^2}{(1+r^2)(1+r_1^2)(r^2+r_1^2)}.
\label{eq:br21}
\ee

The result is

\be
\epsilon=0.22269,
\label{eq:br22}
\ee
a value which has to be compared with Eq. (\ref{eq:br11}).
The corresponding Odderon intercept is
\be
\alpha_O(0)= 0.96.
\label{eq:br23}
\ee

By comparing the values (24) and (14), one sees that there is only a 2\%
difference between the ``exact'' result and the variational one.

We draw from this section the conclusion that the Odderon intercept is very
close to 1, i.e. is much higher than the 1/2 ($\rho$,$\omega$) 
intercept. We therefore expect important Odderon effects at high energy.

The LLA result shows that the gap $\alpha_{O}(0)-1$ is surprisingly small 
and therefore very sensitive to higher order corrections. The crucial 
problem of knowing if $\alpha_{O}(0)$ is bigger than 1, smaller than 1 or 
just equal to one, is therefore still an open problem. The last case 
in the one I favor. The best way of solving this problem is, in my 
opinion, the study of the non-perturbative Odderon. Promising results 
on this line were already obtained\cite{De99}.

Let me add, before closing this section, that very recently, G.P. 
Korchemsky and J. Wosiek obtained a new representation for the 
Odderon wave function\cite{Ko99}, which is in agreement with the 
results of Refs. 7 and 8 and which allows identification of a new quantum 
number - \textit{triality} - associated with the 
Odderon.

An intriguing problem is if there is or not a Pomeron-Odderon 
exchange-degeneracy, like in the non-leading reggeon 
$\rho-\omega-f-A_{2}$ sector. The study of the $C$-even state of 3 
reggeized gluons is crucial in this context\cite{LiPr}.

\section{The Odderon phenomenology}
\subsection{What is the problem with the Odderon ?}

A quarter of century after its birth the Odderon has still an 
uncertain existential status. Of course, there are some interesting experimental 
indications of its existence : 
\begin{itemize}
        \item  [-]
the difference between $pp$ and $\bar pp$ differential cross-sections 
in the dip-shoulder region at ISR energies\cite{Br85} ;
        \item  [-]
the unusual shape of the polarisation in $\pi^-p\to\pi^on$ at low 
energies\cite{Ap82} (indicating a $\rho$-type Odderon, distinct from 
the $\omega$-type Odderon as required in LLA) ;
        \item  [-]
the extraction of the semi-theoretical $\rho$-parameter from the 
$dN/dt$ UA4/2 $\bar pp$ data in the presence of oscillations at very 
small $t$ and high energies or of a more complicated phase of the 
forward scattering amplitude\cite{UA493}.
\end{itemize}

However, these experimental indications are either isolated or 
controversial.

The real problem with the Odderon is the paradoxical scarcity of the 
high-energy data in hadron-hadron scattering, leading to an excessive 
focus on the $pp$ and $\bar pp$ scattering. In other words we fit 
high-energy parameters by using mainly low-energy data and, even 
worse, we draw conclusions about the Odderon based only on $pp$ and 
$\bar pp$ scattering. The folklore about the "suppression" of the 
Odderon has its source in these facts. The recent shift in attention 
from $pp$ and $\bar pp$ scattering towards $ep$ scattering is, in this 
context, very positive.

\subsection{HERA-Odderon phenomenology}
The most active team in the Odderon phenomenology in the last two 
years is, of course, the Heidelberg group (H.G. Dosch, O. Nachtmann, 
E.R. Berger) and its associates (A. Donnachie, P.V. Landshoff, W. 
Kilian, M. Rueter). In a series of papers\cite{Ki98},\cite{Ru96}, 
they produced impressive results with a solid theoretical ground.

For example, the pseudoscalar meson production
\begin{equation}
        e^\pm p\to e^\pm pPS, \mbox{ where }PS=\pi^{o},\ \eta,\ \eta',\ 
        \eta_{c}
        \label{25}
\end{equation}
at HERA ($\sqrt{s}=300.6$ GeV) constitutes a direct probe for the 
Odderon. The Odderon is here in competition with the photon only (see 
Fig. 2) : the Odderon contribution is not obscured by the huge Pomeron 
contribution as in the hadron-hadron reactions.

By taking, as a toy model, the Odderon as a Regge pole located near 
$J=1$,
\begin{equation}
        \eta_{O}\beta(t)
        \left(
        \frac{s}{s_{0}}
        \right)^{\alpha_{O}(t)-1}
        \xi(t)
        \label{26}
\end{equation}
where $\alpha_{O}(0)\simeq 1$, $\xi(t)$ is the signature fractor, 
$\beta\vert t\vert$-the Odderon residue and $\eta_{O}=\pm 1$ (due to 
the absence of a positivity property for the Odderon contribution), 
one gets important Odderon effects in a variety of observables. I 
give just one example in Fig. 3 : the $p_{\perp}$ distribution for 
pion production in the photoproduction region. One can see from Fig. 3 
(where $c_O$ is proportional to $\eta_O$) that dramatic effects are induced
 by the presence of the Odderon. 
Moreover, one can see that the effects for the case $\eta_{O}=+1$ are 
drastically different from the ones for $\eta_{O}=-1$ : the HERA data 
can give important indications on the sign of $\Delta\sigma$ - the 
difference between hadron-hadron and hadron-antihadron total 
cross-sections.

An important theoretical ingredient in several works of the Heidelberg 
group\cite{Ru96} is the Stochastic Vacuum Model (SVM)\cite{Do88} which 
established a very interesting connection between the topological 
Y-shape of the baryons and the coupling of the Odderon. If the angle 
between two sheets of the baryon is very small (i.e. the baryon has a 
diquark-quark structure) the Regge-pole Odderon is suppressed. An 
interesting process is the photoproduction of pions with single 
dissociation (breakup of the target proton), because this process is 
independent of the particular structure of the baryon. Its cross-section is
$\simeq 300$ nb, i.e. 50 times larger for the corresponding one in 
the elastic photoproduction.

Of course, taking the Odderon as a Regge-pole could be too 
simplistic : for example, perturbative QCD indicates a more complicate 
singularity structure in the complex $J$-plane.
Moreover, we showed longtime ago\cite{GLN90} that the Regge-pole 
Odderon induces an overall shift of the low-energy data for the real 
parts which are already very well described by the Pomeron and the 
secondary Regge poles. Therefore, the Regge-pole Odderon is the worst 
case to be considered as a possible Odderon singularity. However, as 
a toy model, it can be still used as an illustration of typical 
Odderon effects, present even if the Odderon coupling is very much 
suppressed as compared with the Pomeron one.

A very interesting Odderon effect was recently studied by S.J. 
Brodsky et al.\cite{Br99} : the assymetry in the fractional energy of 
charm versus anticharm jets in the diffractive photoproduction
$\gamma p\to c\bar cY$ at HERA. This assymmetry is very sensitive to 
the Pomeron-Odderon interference (Fig. 4) : it measures the Odderon 
amplitude linearly. Namely
\begin{equation}
        A(t,M^{2}_{x},z_{c})\simeq
        \epsilon
        \frac{\sin[(\pi/2)(\alpha_{O}-\alpha_{P})]}{\cos((\pi/2)\alpha_{O})}
        \left(
        \frac{s_{\gamma p}}{M^{2}_{x}}
        \right)^{\alpha_{O}-\alpha_{P}}
        \label{27}
\end{equation}
where $z_{c}=E_{c(\bar c)}/E_{\gamma}$ and $\epsilon=\pm 1$. It can 
be seen from eq.(\ref{27}) that the sign of the asymmetry is 
controlled by the gap ($\alpha_{O}-1$) and the energy dependence by 
the gap $(\alpha_{O}-\alpha_{P})$. By taking, as a numerical 
illustration, $\alpha_{P}=1.13$ and $\alpha_{O}=0.95$, Brodsky et al. 
get a lower bound for the asymmetry equal to 15\%.

Before closing this section, let me mention other interesting 
phenomenological studies of the Odderon : diffractive $C=+$ neutral 
meson production from virtual photons\cite{En98}, exclusive 
$\eta_{c}$ photo- and electroproduction\cite{Cz97}, exclusive $f_{2}$ 
leptoproduction\cite{Ry98} and single-spin asymmetries for 
small-angle pion production in hadron colisions\cite{Ah99}.

\section{Conclusions : the Odderon in the future}
As we understood from the talk of S. Weisz at this 
Conference\cite{To99}, the TOTEM experiment at LHC will not be very 
helpful for the Odderon physics : at least in the initial stage of 
the experiment, priority will be given to just mesuring the $pp$ 
total cross-section.

Great hopes arise from the talk of S.B. Nurushev\cite{Nuru} concerning 
the R7 experiment at RHIC\cite{R795} with a higher luminosity that the 
UA4/2 experiment at CERN. Several measurements concern directly the 
Odderon physics : extraction of the $\rho$ parameter with a 
precision $\delta\rho=0.01$ ; research for oscilation at very 
small $t$ ($\vert t\vert\simeq (1-4)\cdot 10^{-4}\ \mbox{GeV}^{2}$ ; 
the evolution of the dip in $d\sigma/dt$ ; the polarisation parameters. 
In connection with this last measurement, E. Leader and T.L. Trueman 
showed in a very recent paper\cite{Le99} the high sensitivity of the 
spin dependence of $pp$ scattering, particularly the parameter 
$A_{NN}$, to the Odderon. 

We had a nice surprise learning that the 
$\bar pp$ option at RHIC is realistic. That would allow the detection 
of the Odderon through the measurement of the difference 
$\Delta\sigma$ of the $\bar pp$ and $pp$ total cross-sections : the 
Regge-pole model predicts $\Delta\sigma=40\ \mu b$ at 
RHIC while the maximal-Odderon approach\cite{Lu73} predicts\cite{Br85}
$\Delta\sigma=-2.4$ mb. The expected precision at RHIC being 
$\Delta\sigma=0.5$ mb, one can clearly detect the presence of the 
Odderon through $\Delta\sigma$. 

Let me also stress the importance of 
the systematic study of the energy-dependence of $\sigma_{T}(s)$ in 
the RHIC range 50 GeV $\leq\sqrt{s}\leq$ 500 GeV. A huge gap in the 
high-energy hadron data will be filled.

However, we have not to wait till the next millenium in order to get a 
long awaited evidence of the Odderon. The results of the H$_{1}$ 
experiment on the pseudoscalar meson production at HERA\cite{H197} 
will be soon available.

\section*{Acknowledgments}
It is a great pleasure to thank Prof. Vladimir Petrov and his young 
and dynamic staff for the remarkable organization of this Conference.

\newpage
\begin{center}
\includegraphics*[scale=0.7]{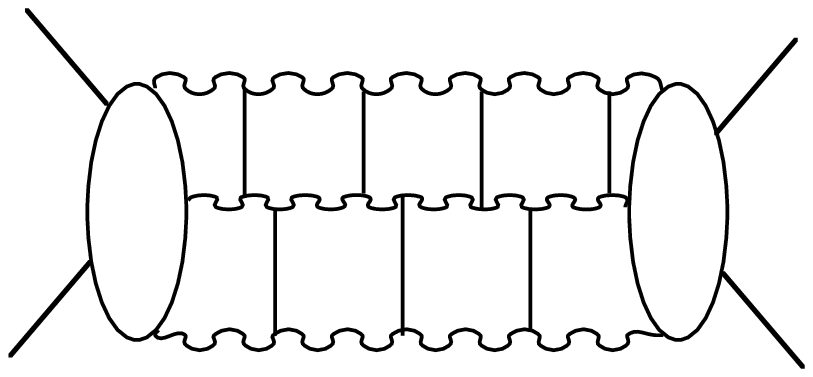}
\vspace*{0.3cm}

Fig. 1. The Odderon in QCD as a C-odd state of 3 reggeized gluons.
\end{center}


\begin{minipage}[c]{.40\linewidth}\hfill
\begin{center}
\includegraphics*[scale=0.6]{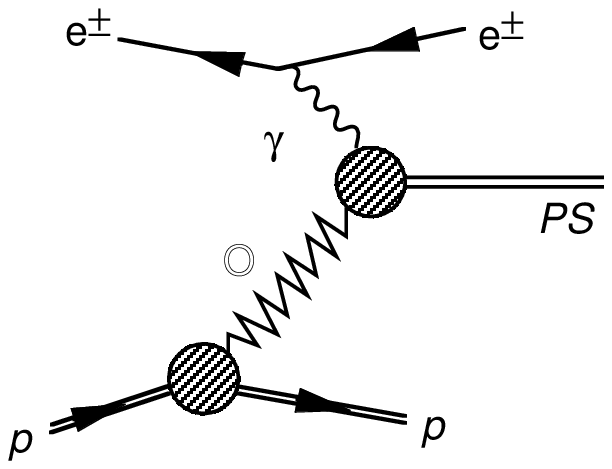}
\vspace*{0.3cm}

Fig. 2. The Odderon exchange in $e^{\pm} p\to e^{\pm}pPS$.
\end{center}
\end{minipage}
\begin{minipage}[c]{.55\linewidth}
\begin{center}
\vspace*{0.9cm}
\includegraphics*[scale=0.62]{FIG3}
\vspace*{0.5cm}

Fig. 3. The $p_{\perp}$ distribution for pion production in the 
photoproduction region (Ref. 17).
\end{center}
\end{minipage}

\vspace*{1cm}
\begin{center}
\includegraphics*[scale=0.5]{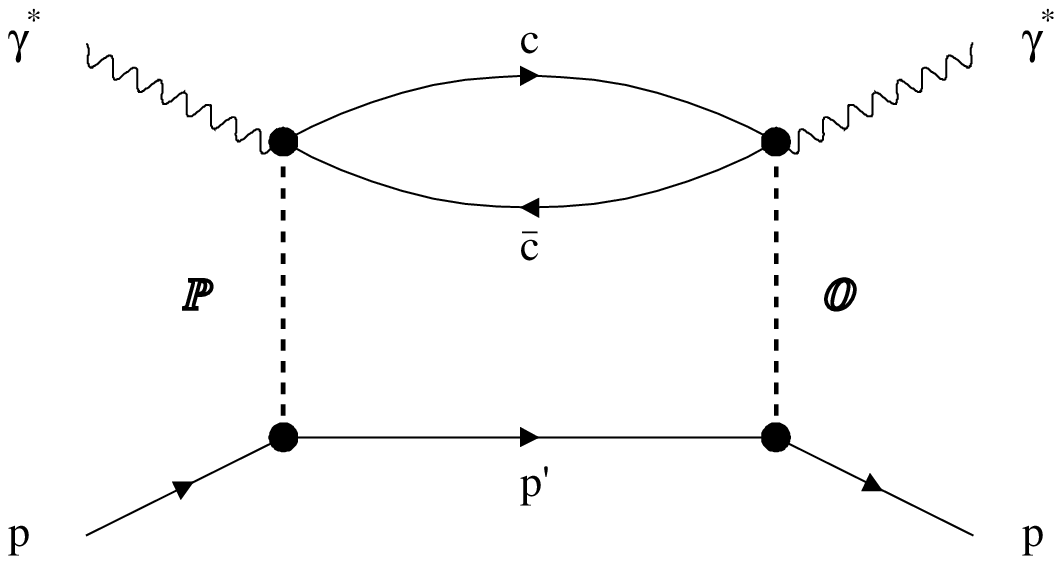}
\vspace*{0.6cm}

Fig. 4. The Pomeron-Odderon interference in $\gamma p\to c\bar c p'$
(Ref. 21).
\end{center}

\end{document}